*Earth and Planetary Science* | Volume 02 | Issue 02 | October 2023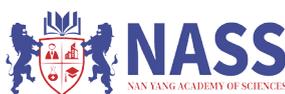

**Earth and Planetary Science**

https://journals.nasspublishing.com/index.php/eps

## REVIEW
# Origin of the Moon and Lunar Water

Nick Gorkavyi[*]

Science Systems and Applications, Inc., Lanham, MD 20706, USA
ARTICLE INFO

*Article history*
Received: 31 August 2023
Revised: 11 October 2023
Accepted: 16 October 2023
Published Online: 20 October 2023

*Keywords*:
Moon
Mega-impact
Water
Asteroid satellites
ABSTRACT

Three principal concepts regarding lunar formation have been examined: the accretion hypothesis, the mega-impact theory, and the multi-impact model. The multi-impact model amalgamates the salient facets of the mega-impact theory and the accretion hypothesis. As per this model, fragments of the terrestrial crust are ejected into space during collisions with numerous planetesimals (proto-asteroids) with diameters around 10-100 kilometers. In the vicinity of Earth's orbit, this ejecta interacts with the accretion disk, augmenting its mass. Numerical computations demonstrate that particles within the prograde-rotating accretion disk effectively capture prograde-rotating ejecta while shedding retrograde-rotating ejecta onto the planetary body. The multi-impact theory provides an explanation for the creation of not just the Moon and Charon but also the satellites of asteroids. Several predictions regarding the parameters of asteroid satellites posited by this theory have already been validated through statistical analysis of binary asteroids. Different models of lunar formation yield varied conclusions regarding the quantity of lunar water, its subsurface distribution, and isotopic composition. The mega-impact theory postulates the genesis of a largely desiccated Moon. Consequently, the modest lunar water content may arise from comets, solar wind, or the transport of water from Earth's atmosphere. These mechanisms for lunar water formation imply its superficial occurrence in polar regions and substantial deviation of the deuterium-to-hydrogen (D/H) ratio from terrestrial norms. Conversely, the multi-impact theory posits lunar water's origin from planetesimals, akin to terrestrial water. Thus, a significant quantity of lunar water is inferred, expected to be pervasive across the lunar surface at varying depths, and possessing isotopic composition analogous to terrestrial water. Geomorphological structures in the lunar polar regions (smoothed craters, landslides, regular patterns) suggest the presence of a substantial permafrost layer with an approximate thickness of a kilometer.
## 1. Introduction

In recent years, various lunar exploration programs have been actively advancing. The most ambitious among them are the programs of the United States [1] and China, yet Russia, India, and Japan also have their lunar initiatives. As part of the Indian project "Chandrayaan-3", on August 23, 2023, a lander along with a mobile rover successfully touched down in the vicinity of the Moon's southern pole. The "Smart Lander for Investigating Moon" (SLIM), developed in Japan, was launched to the Moon on September 6, 2023. In November 2024, the American lunar rover "VIPER" (Volatiles Investigating Polar Exploration Rover) is planned to touch down in the vicinity of the Moon's southern pole. One of the primary objectives facing these lunar landers is the search for and study of lunar water. The investigation of lunar water sheds light on the Moon's origin, making it intriguing to analyze the lunar water predictions of different lunar origin models. Early descriptions of lunar formation models (such as capture and fission due to centrifugal force) can be found in Ringwood's book [2]. Herein, our focus centers on three pivotal theories of lunar formation: the accretion model, the mega-impact theory, and the multi-impact model.

*Corresponding Author:
Nick Gorkavyi,
Science Systems and Applications, Inc., Lanham, MD 20706, USA;
*Email: nick.gorkavyi@ssaihq.com*
DOI: https://doi.org/10.36956/eps.v2i2.940Copyright © 2023 by the author(s). Published by Nan Yang Academy of Sciences Pte. Ltd. This is an open access article under the Creative Commons Attribution-NonCommercial 4.0 International (CC BY-NC 4.0) License. (https://creativecommons.org/licenses/by-nc/4.0/).86

## 2. Accretion Model of Lunar Formation and Its Challenges

The Safronov group successfully formulated an accretion theory for the formation of terrestrial planets [3]. Ruskol proposed a similar accretion theory for the Moon's formation from near-Earth protosatellite swarms [4]. Such a swarm around Earth was envisioned to accumulate from particles moving along heliocentric orbits, colliding within Earth's gravitational field, and transitioning to near-Earth orbits, thus forming a protosatellite swarm or disk. The accretion model faces two significant challenges:

a. The accretion mechanism leads to a relatively low mass fraction ($10^{-3}$ to $10^{-4}$ of the planet's mass) for the protosatellite swarm, which is inadequate to account for the Moon's substantial mass fraction (1/81 or 1.2% of Earth's mass, a record in the Solar System as of 1975);

b. There is no explanation for the Moon's average chemical composition, including its low density (3.3 g/cm$^3$) and iron deficiency (13% FeO) [5]. If the accretion theory holds, the Moon, formed similarly to Earth from planetesimals, should exhibit analogous parameters: a density of 5.5 g/cm$^3$ and an average iron content of 31% FeO [6].

## 3. Mega-Impact Model of Formation and Its Challenges

Until 1610, the Moon was perceived as a singular celestial body; no other natural satellites were known within the Solar System. However, four centuries ago, Galilei's observations revealed the presence of four new moons around Jupiter, marking the initiation of a new era of Solar System exploration. In terms of relative mass, the Moon exceeded all previously known satellites before 1978.

The Moon's distinct status as a one-of-a-kind natural satellite led to the development of a unique theory about its origin: the mega-impact theory. This theory was initially proposed by R. Daly in 1946 and was further detailed in a 1975 publication by W. Hartmann and D. Davis [7]. According to this theory, larger celestial bodies that were forming near Earth's orbit could have collided with Earth during the first $10^7$-$10^8$ years, approximately 4.5 billion years ago. If a planetesimal with a radius of about 1200 km struck Earth's surface at a speed of 13 km/sec, it could have provided enough kinetic energy to eject two lunar masses at nearly escape velocities [7]. Hartmann and Davis concluded that this model could explain the Moon's characteristics, such as its reduced iron content, increased abundance of refractory elements, and depletion of volatile substances. In accordance with the Hartmann-Davis hypothesis, the protosatellite disk, and subsequently the moon, are postulated to form from ejected fragments of Earth's crust and upper mantle, possessing a density similar to that of the Moon, approximately 3 g/cm$^3$. This explanation also addresses the Moon's iron deficiency, as Earth's crust and mantle exhibit diminished iron content (6% FeO) [6] due to its concentration within the molten Earth's core, where FeO constitutes 83% [6].

Hartmann and Davis employed a very moderate impactor mass estimate of $6.7 \times 10^{-3}$ of Earth's mass, merely half the mass of the Moon, for their calculations. However, Hartmann and Davis's qualitative model did not account for the intricate issue of retaining material in near-Earth orbit. Within a two-body problem context, matter ejected from Earth due to impact cannot persist in near-Earth orbit. This ejecta must acquire hyperbolic velocity and escape onto heliocentric orbits or enter elliptical orbits around Earth, returning to Earth's surface within a single revolution due to the lower part of such an orbit intersecting Earth's surface.

Consequently, when scientists embarked on detailed calculations, they significantly deviated from the original Hartmann and Davis estimates. They augmented the impactor mass to the extent that the dynamics of smaller fragments conformed to the three-body problem, as fragments moved between Earth and the surviving part of the impactor. This adjusted model presented the possibility of retaining at least a fraction of the ejected material in orbit.

In accordance with this model, there was a colossal collision around 4.5 billion years ago involving Earth and a massive celestial body named Theia. Theia was approx-imately the size of Mars, with a diameter of around 7,000 km. This extraordinary event resulted in fragments from the collision being placed into an orbit around Earth.

In the most contemporary calculations, the mass of the impacting planet has reached approximately half of the proto-Earth's mass [8]. As a result of this catastrophic impact around Earth, a protolunar disk with significant angular momentum emerged. The model featuring a massive impactor resolved the issue of retaining a small amount of ejected material in near-Earth orbit but gave rise to another predicament: it was revealed that, under such a scenario, the Moon is formed not only from Earth's crust and mantle but also from material derived from the impacting Theia. Since Theia's isotopic composition is expected to differ from that of Earth, this implies that the isotopic composition of the Moon, which comprises a substantial portion of Theia's material, must also differ from that of Earth. However, as demonstrated in a series of studies, the isotopic composition of oxygen, titanium, and tungsten in lunar and terrestrial rocks aligns entirely.

The uniformity of isotopes implies that the material found on the Moon originated from the mantle of the early Earth [9].

There are additional arguments against the mega-impact model. Geochemical analyses have led to the following conclusions regarding the catastrophic impact model [5]:

a. The Moon, which originated from Earth, must be younger than Earth;

b. Due to the mega-impact, both the Moon and Earth should have experienced melting. This suggests that both Earth and the Moon likely had oceans of molten magma;

c. The Moon's lack of volatile elements can be attributed to the heating of Earth's material ejected into orbit during the mega-impact.

However, empirical data from these geochemical analyses provide a somewhat different perspective [5]:

a. The Moon is older than Earth or, to be more precise, the Moon's core may have formed before Earth's core;

b. The Moon appears to have been relatively cool and might not have been entirely covered in molten magma. Furthermore, geochemical data contradicts the idea of a molten mantle ocean on Earth. For instance, the Earth's current mantle shows less differentiation than it would if there had been a molten magma ocean in the distant past;

c. The composition of volatile elements found on the Moon doesn't align with the mega-impact theory. These elements cannot be explained as originating from Earth's mantle through heating processes.

As a result, in the year 2000, geochemists J.H. Jones and H. Palme supported the model of lunar formation from a near-Earth protosatellite disk generated through accretion [5].

Charon, one of Pluto's moons, was discovered in 1978. Despite Pluto being a small, cold dwarf planet on the outskirts of the Solar System, Charon is relatively massive compared to its parent body, accounting for 12 percent of Pluto's mass. In contrast, the Moon only makes up 1.2 percent of Earth's mass, thereby diminishing the Moon's status as a unique satellite in terms of its relative mass. Furthermore, the phenomenon of substantial (approximately 10%) duplicity was observed among small solid-surfaced celestial bodies—asteroids and trans-Neptunian objects [10]. In 1994, the first satellite of an asteroid (Ida) was photographed, followed by the discovery of the first triple asteroid (Sylvia) in 2005. Additionally, apart from Charon, four smaller and more distant satellites were detected around Pluto. It is noteworthy that the origin of Charon and asteroidal satellites within the framework of the mega-impact model appears highly implausible. This is due to the fact that collisions among bodies in the asteroid belt occur at speeds of several kilometers per second, while the initial escape velocity for asteroids is lower by 2-3 orders of magnitude. The matter ejected by a massive impactor is simply unfeasible to retain in orbit around an asteroid with weak gravity.

In 1989, Ringwood published an article [11] in which he noted that the current version of this hypothesis is unlikely to be true for several reasons:

1) The probability of such an event is low.

2) If such a collision had occurred, it would have melted and separated the Earth's materials, resulting in chemical compositions different from what we observe.

3) The mechanics of this collision suggest that the Moon's composition would primarily come from the mantle of the impacting body, while the evidence from geochemistry strongly suggests that most of the Moon's material originated from the Earth's mantle instead.

Ringwood concludes that in the Moon's creation, impactors with masses of 0.001 to 0.01 times that of Earth were involved.

As far back as 1986, Ruskol introduced the concept that the Moon didn't originate from a single massive impact but rather from a series of "macro-impacts", which are essentially the same phenomenon but on a somewhat smaller scale. In these impactful events involving significant celestial bodies, the developing Earth isn't entirely obliterated. Instead, the ejections from the impact craters in the Earth's mantle are so forceful that they partially escape Earth's gravitational pull and spread into the nearby space, forming a swarm of materials [12].

## 4. The Multi-Impact Model and Its Predictions

We maintain that the Hartmann and Davis model [5] was accurate in its assessments of the maximum impactor mass. These assessments aligned with calculations by the Safronov group, which indicated that the mass of the second-largest body within Earth's vicinity should be close to 1% of Earth's mass. The issue of retaining material in near-Earth orbit, which remained unresolved in the Hartmann-Davis model, needed to be addressed not by doubling the impactor mass, but by considering an accretion disk that would inevitably accumulate near Earth. The interaction between the propelled ejecta and particles of the circumplanetary disk altered the trajectory of the ejecta, taking its dynamics beyond the scope of the two-body problem.

In 2004, Gorkavyi proposed the multi-impact model for the formation of celestial bodies like the Moon, Charon, and binary asteroids, which doesn't involve catastrophic events [13]. Here are the key elements of this new model:

**1) Moon Material from Earth**: The majority of the Moon's material came from the Earth's mantle through multiple impacts by large asteroids (approximately 1-100 km in size). This explains why the Moon has a "low-iron" composition, similar to the single-impact model proposed by Hartmann and Davis.

**2) Prograde Protosatellite Disk**: Initially, there was a low-mass prograde protosatellite disk orbiting around the proto-Earth, similar to those seen around other planets.

**3) Debris Collection**: Collisions between Earth debris and particles in the prograde protosatellite disk played a crucial role in collecting the debris into stable orbits. Prograde Earth debris effectively joined the prograde protosatellite disk, while retrograde debris returned to Earth. Due to the prograde rotation of the Earth, the volume of prograde debris exceeds the volume of retrograde ejecta.

**4) Optimal Ring Radius**: Calculations regarding the transfer of angular momentum showed that the protosatellite ring should have an optimal radius, close to the average semi-major axis of debris orbits (A.M. Fridman, N.N. Gorkavyi "Physics of Planetary Rings", Springer, 1999, p.222-233). Earth debris pushed away a smaller ring and decreased the angular momentum of a larger ring.

**5) Formation of Massive Ring**: A massive, low-iron ring near Earth eventually accreted onto the Moon. In the final stages, Earth debris bombarded the newly formed Moon, resulting in a dichotomy of crater populations.

Predictions from this new multi-impact model include:

**a. Satellite Orbits**: Most satellites of asteroids should have prograde and circular orbits that are close to the equator of the central body, similar to the Moon-Earth and Charon-Pluto systems.

**b. Asteroid Rotation**: Asteroids with satellites should rotate faster than single asteroids.

**c. Satellite Absence**: Low-rotating asteroids (or Earth-like planets) should generally lack satellites, as is the observed case with Venus and Mercury.

**d. Satellite Mass**: All else being equal, the relative mass of satellites can be larger for smaller asteroids.

**e. Crater Dichotomy**: The dichotomy of craters should be a common feature of Charon, Pluto, and large asteroid satellites.

**f. Moon Signatures**: The Moon should exhibit chemical, isotopic, and geological signatures of many different asteroid impacts from Earth.

This model provides an alternative explanation for the formation of these celestial bodies without the need for catastrophic events.

This qualitative hypothesis needs to be substantiated through detailed calculations. The most challenging issue pertains to the stability of the protosatellite disk, which is bombarded by ejecta propelled from Earth's surface by impacts from planetesimals. These impacts expel ejecta onto both prograde and retrograde orbits. Neglecting Earth's rotation, the cumulative angular momentum of the ejecta is zero. How does a low-mass protosatellite swarm with positive angular momentum interact with an intense stream of zero-net-momentum ejecta? Can the ejecta precipitate the protosatellite swarm to Earth?

To address these inquiries, a numerical model was developed, which investigated the interaction between protosatellite disk particles and bodies ejected from the Earth's surface [14].

The article by Gorkavyi [14] addresses the planar two-body problem in the case of Earth and a particle under the following conditions (neglecting the influence of the Sun and Earth's rotation effects):

1) The elliptical trajectory of the ejecta particle intersects the circular orbit of the satellite particle in the ascending branch (point A) and descending branch (point B).

2) The major axis and eccentricity of the intersecting orbits are specified, enabling us to readily determine the azimuthal and radial components of velocity for each particle at points A and B.

3) The collision of the two particles forms a cloud of debris, the center of mass of which moves away from the collision point with velocities determined by the conservation of momentum and the mass ratio between the ejecta and satellite particles.

4) With knowledge of the velocity vector of the center of mass of the debris, we determine the major axis and eccentricity of its trajectory and analyze:

a. How the debris orbit changes based on the mass ratio of particles, the orbit parameters of the ejecta, and the radius of the protosatellite swarm particle's orbit.

b. Whether the debris impacts the planet's surface or remains in a stable satellite orbit.

c. Whether the protolunar disk's mass decreases or increases under the influence of planet-originated ejecta. Does the disk maintain its orbital stability?

We postulate that the distribution of ejecta is symmetric with respect to the direction of Earth's revolution—meaning that for a certain number of particles ejected from the planet's surface onto prograde near-Earth orbits, an equal number of particles are ejected onto retrograde orbits. Two primary scenarios exist for the interaction between the disk particles and ejecta:

1) Ejecta with retrograde orbital rotation interacts with particles from the prograde disk, thus inevitably impacting Earth and carrying along a disk particle, if its mass is smaller or comparable to that of the ejecta particle.

2) Ejecta with prograde orbital rotation interacts with particles from the prograde disk and is highly likely to transition to a stable satellite orbit. Even a slight perturbation from a disk particle is sufficient to induce an increase in the ejecta's perigee, thus avoiding collision with Earth.

Figure 1 illustrates the growth and degradation profiles of a disk extending above the Earth's surface by 10,000 km. A study was conducted involving 1000 distinct values for ejecta particle masses ranging from 0.1 to 10 relative to the mass of disk particles. The number of particles is inversely proportional to their mass. Additionally, 1000 disk orbits (with a step of 10 km) of varying radii and 400 ejecta orbits with uniformly distributed eccentricities from nearly circular to 1 were examined. All ejecta orbits are tangential to the Earth's surface, thus their semi-major axes are consistent with this condition. The probability of ejecta interaction with each disk orbit was conventionally set at 1%. It is assumed that the debris cloud contributes to the disk orbit with a radius equal to the semi-major axis of the debris cloud orbit. The debris cloud distribution was examined within a space twice the size of the disk, up to 20,000 km above the Earth's surface. A total of 400 million debris cloud trajectories were analyzed.

It is evident that the parts of the protosatellite disk closest to the planet are vanishing (due to particle outward migration or collision with the planet), while in the rest of the disk, a confident growth (gain) is observed, dominating over the reduction (loss). The disk's growth is attributed to the accretion of prograde ejecta, which compensates for the disk's reduction caused by bombardment from retrograde ejecta. The examined mechanism effectively selects ejecta trajectories that align with the pro-lunar disk's orbital motion. The qualitative results remain unchanged under substantial variations in the size distribution law of the ejecta particles or when considering an extensive disk extending up to 100,000 kilometers from the planet's surface. These calculations compellingly demonstrate the dynamic efficiency of the discussed lunar formation mechanism.

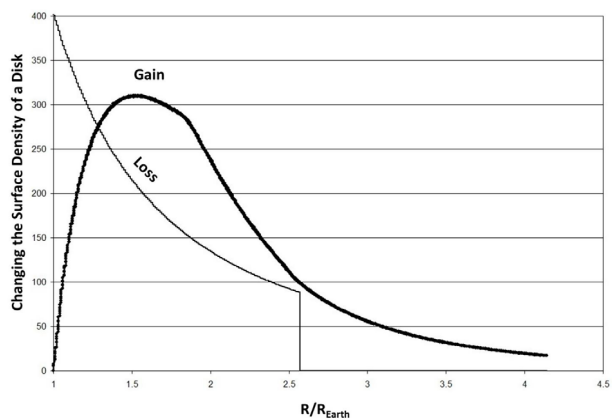

**Figure 1.** (From the article [14]). This graph illustrates the changes in surface density within the protosatellite disk caused by its interaction with the material ejected from Earth, with values measured in arbitrary units. Initially, the disk, which extends 10,000 kilometers above Earth's surface, is considered to be uniform in composition. The slender curve shows how the surface density of the disk decreases over time due to the material being expelled onto Earth after interactions with both prograde and retrograde ejecta trajectories. On the other hand, the bold curve depicts how the surface density of the disk increases over time as a result of material accreting onto it from the direct ejecta. The most favorable conditions for satellite formation occur within the zone of 2-3 Earth radii, where the increase (gain) in disk surface density significantly surpasses the losses.

In 2017, Rufu, Aharanson, and Perets published a similar article in Nature-Geoscience on the multi-impact formation of the Moon, demonstrating that the Moon's origin could be attributed to numerous moderate impacts rather than a single mega-impact event [15]. In their study, they examined a sequence of 20 impacts from bodies with masses ranging from 0.1 to 0.01 times that of Earth, which generated disks and subsequently formed satellites. These 20 satellites eventually merged to form a single larger Moon. The structure of the protosatellite disk presented in the Rufu, Aharanson, and Perets article [15] aligns well with Figure 1 from the article [14].

The lunar formation model, which combines the existence of an accretion disk around Earth along with the continuous replenishment of this disk through multiple macro-impact events, is actively developed within the Moscow Safronov group. See, for instance, the work by Adushkin et al. [16] and the references therein. Ruskol discusses the asymmetry of craters and maria on the visible and far sides of the Moon [4]. The higher concentration of lunar maria (dark, flat plains) on the side of the Moon that faces Earth, particularly in the leading quarter of the lunar disk, could suggest that the Moon has been impacted by material ejected from Earth. Thus, the "maria" asymmetry suggests the occurrence of numerous and temporally distributed strong asteroid impacts, transporting material from Earth to the Moon.

Hence, the multi-impact theory of lunar formation combines the strengths of both the accretion theory and the original Hartmann-Davis mega-impact model. This model is independently developed by three groups of researchers, yielding similar outcomes. It remains consistent with contemporary geochemical data and avoids dynamic issues. A significant advantage of the multi-impact theory lies in its applicability not only to the Moon's formation but also to the formation of Charon and asteroid satellites [17]. This broadens the opportunities for testing the theory against a substantial volume of observational data.

## 5. Comparison of Moon Formation and Binary Asteroids

In the study by Gaftonyuk and Gorkavyi, a database containing 113 binary asteroids was examined, with rotational periods and diameters of the primary components determined [18]. The predictions of the multi-impact model [13,14,17] were fully corroborated. The average rotation rate of single asteroids amounts to 3.54 rotations per day, whereas the average rotation rate of the primary bodies of binary asteroids was found to be nearly twice as high: 6.56 ± 0.28 rotations per day.

Furthermore, it has also been demonstrated that:

1) The satellites of asteroids exhibit solely direct orbits, typically characterized by low eccentricities and inclinations relative to the equator of the central body, thereby falling within the class of regular satellites.

2) The duality of asteroids (percentage of bodies with satellites) rapidly increases with the rotational speed of the primary body.

3) The percentage of duality among asteroids depends on the size of the main asteroid, and there is a minimum number of binary asteroids found among those sized between 10-100 km. Asteroids with satellites are most commonly observed in two size ranges: those smaller than 10 km and those larger than 100 km.

Consequently, asteroid satellite systems are regular and akin to satellite systems of typical planets. When considering mass relative to their respective primary bodies, the Moon is positioned somewhere in the middle of the distribution of asteroid satellites, thereby definitively shedding the aura of uniqueness. Models of Moon formation in proximity to Earth and satellites around solid-surfaced planets must be unified, elucidating both the Moon and asteroid satellites. A potential question might arise: if the mechanism of asteroid satellite formation lacks substantial constraints on its efficacy, then why do only about 10% of asteroids manifest as binaries or possess satellites?

We posit that the actual percentage of duality among asteroids and trans-Neptunian objects is considerably higher; however, when the protosatellite disk transforms into an actual satellite, the mechanism of selection exclusively by direct ejection ceases to operate. A substantial satellite absorbs ejected material with a net zero angular momentum and converges with the primary body. The orbital velocity of the satellite is contingent upon the mass of the central body. Depending on the rotational velocity of the central body, the satellite can gently merge with it, thereby forming a binary system. Notably, certain asteroids, such as 486958 Arrokoth, exhibit a dumbbell shape or that of two agglomerated bodies (Figure 2).

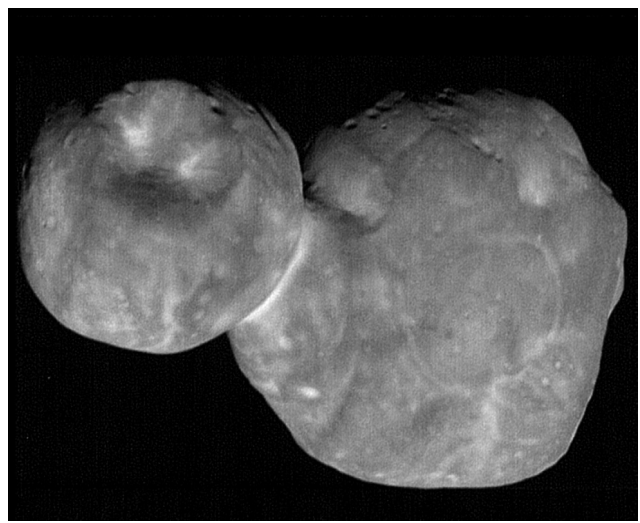

**Figure 2**. The most detailed images of 486958 Arrokoth (Wenu 21.20 × 19.90 × 9.05 km; Weeyo 15.75 × 13.85 × 9.75 km [19]). New Horizons spacecraft, 1 January 2019 (https://www.nasa.gov/feature/new-horizons-spacecraft-returns-its-sharpest-views-of-ultima-thule).

The formation of another satellite on an outer orbit around such a binary system is also possible. A third body, undergoing a similar evolution, may decelerate and become incorporated into the already merged pair of bodies. This appears to be how the triple body of asteroid 9969 Braille was formed.

In the scenario where the moon follows a circular orbit and collides tangentially with a large, quasi-spherical asteroid, it would leave an elongated trench (fossa) on the surface of the primary body. The asteroid Vesta, a large and fast-rotating celestial object with a diameter of 525 km, is an example of this phenomenon. Despite its size, Vesta does not possess any moons. However, a distinctive series of enormous fossae were discovered on its surface (indicated by blue arrows in Figure 3). These fossae are found in two formations: one near the equator (Divalia Fossae Formation) and another at approximately 30°N (Saturnalia Fossae Formation). These grooves extend to lengths of 300-400 km, widths of 10-20 km, and depths of up to 5 km (Figure 3). A moon orbiting Vesta would need to travel at a velocity of 900 km/h. If its orbital rotation aligns with the direction of Vesta's rotation, the moon would move relative to Vesta's surface at approximately 600 km/h. According to calculations, a moon with a mass roughly $10^{-3}$ times that of Vesta, having the same density, would possess enough kinetic energy before it completely breaks apart to create a canyon. This canyon would be 5 kilometers deep, 10 kilometers wide, and 1,000 kilometers long, even if the material it is moving through is as tough as durable granite. The specific energy required for this fragmentation process would be approximately $10^9$ erg/cm$^3$. The presence of two distinct sets of grooves suggests that Vesta encountered and incorporated two moons with inclined orbits. The author of this theory first introduced this concept in an interview with "New Scientist"[20] in 2015 and later elaborated on it in a more comprehensive discussion published in 2019[21].

Attention is drawn to intriguing spiral structures in the polar regions, particularly prominent in the vicinity of the southern pole (indicated by yellow arrows). As calculations reveal, the absorption of a 50-kilometer-diameter satellite, moving in the direction of the blue arrows at a velocity of approximately 200 m/s, should lead not only to the formation of deep valleys, expanding the satellite's energy, but also to a global acceleration of the asteroidal crust in the equatorial region, utilizing the angular momentum of the satellite. Following the satellite's absorption, the crust with a thickness of around 10 km should acquire additional rotation at a rate of about one meter per second relative to the satellite's core and its polar regions. This additional angular momentum may induce a displacement of the crust with respect to the asteroid's main mass by ~100 km, resulting in the appearance of spiral structures in the polar region.

Another scenario that should be considered is a gentler interaction between an asteroid and its satellite. This scenario assumes that the moon has relatively low energy and cannot create a canyon or crater that is larger in volume than the moon itself. In such situations, for the creation of a dumbbell-shaped asteroid, certain conditions must be met. These conditions include a constrained orbital velocity for the moon, represented as V, where $V^2 < 2E_v/\rho$. Here, $E_v$ signifies the specific volumetric energy required for the asteroid's breakup, and $\rho$ represents the moon's density. This velocity threshold is approximately 100 meters per second. Consequently, rapidly rotating asteroids with diameters of several hundred kilometers or less can display equatorial grooves, which essentially serve as the impact scars left by their moons. For asteroids with diameters less than one hundred kilometers, the interaction with a moon should result in the formation of a dumbbell-shaped structure.

In the issue of lunar formation, the primary question revolves around the source of material required for the creation of such a satellite. In the issue of asteroid belt formation, the key query pertains to the whereabouts of 99.97% of the mass of the belt. Interestingly, these two questions are intricately interconnected.

The actual masses of Mars and the asteroid belt differ significantly from the values that would be expected based on theoretical predictions derived from extrapolating the density of solid materials present in the protoplanetary disk near Venus and Earth. During collisions of planetesimals, a significant amount of dust is generated. A more detailed analysis has revealed that 10% of dust particles ranging in size from 0.2 to 40 micrometers acquire a significant eccentricity in their orbits due to the pressure of solar radiation. Eventually, Jupiter's gravitational field scatters these particles, causing them to exit the Solar System, on average, within about 30,000 years[22]. This process could explain a substantial loss of mass in both the asteroid belt and the Martian region[23]. The effectiveness of removing dust particles from asteroids, despite their weak gravitational pull, is supported by observations from spaceborne images. These images show that the surfaces of small asteroids exhibit rocky features, while the larger Moon's surface is covered by fine dust, which remains in place due to the Moon's gravitational forces (Figure 4).

The isotopic composition of noble gases found in surface regolith samples taken from asteroid 25143 Itokawa suggests that over a span of more than one million years, this asteroid, with an average diameter of around 330 meters, experiences a loss of surface material equivalent to a layer several tens of centimeters thick[24].

Hence, asteroids experience a reduction in mass as they lose a significant portion of their substance, while simultaneously generating a region conducive to satellite formation around them. It is the satellites that capture a fraction of the material ejected from the asteroid's surface and thereby grow. Following a similar mechanism, the

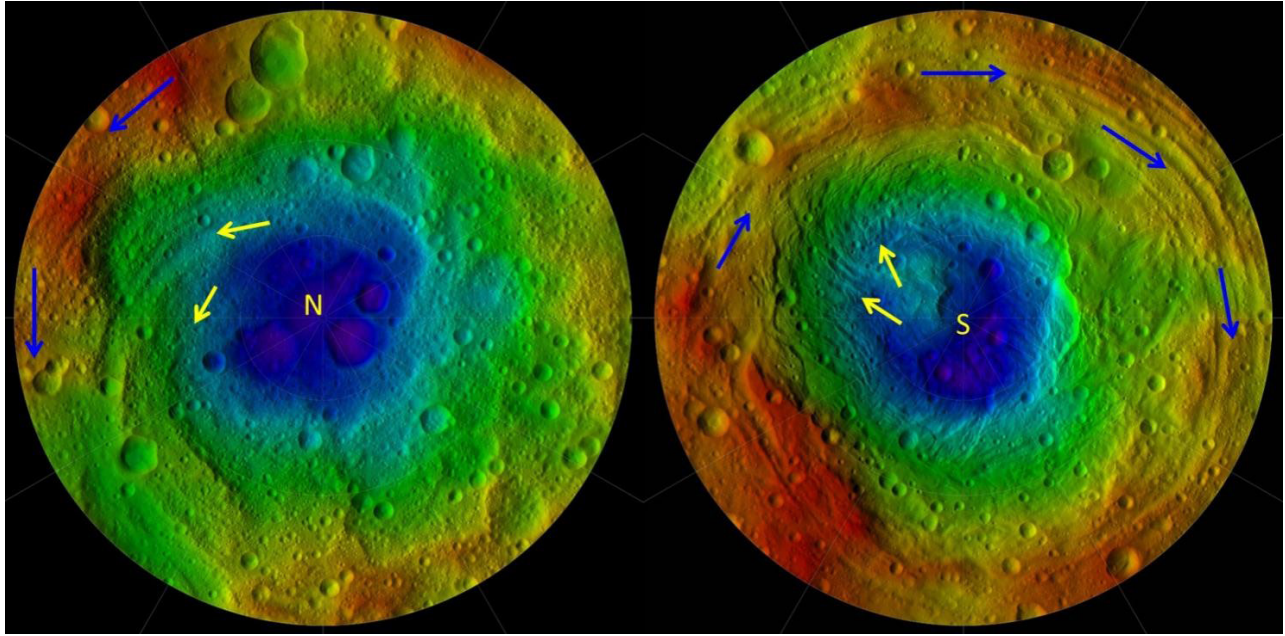

**Figure 3**. The relief map depicts the northern (left) and southern (right) hemispheres of asteroid 4 Vesta. The equatorial region is omitted from the map. The color scheme corresponds to the distance from Vesta's center, with lower elevations represented in violet and higher elevations in red. This map is a result of the Dawn mission and was created using data from NASA, Johns Hopkins Applied Physics Laboratory, Southwest Research Institute, and the National Optical Astronomy Observatory. Blue arrows on the map indicate the direction of the satellite's orbital movement, while yellow inner arrows highlight spiral structures present in the polar regions.

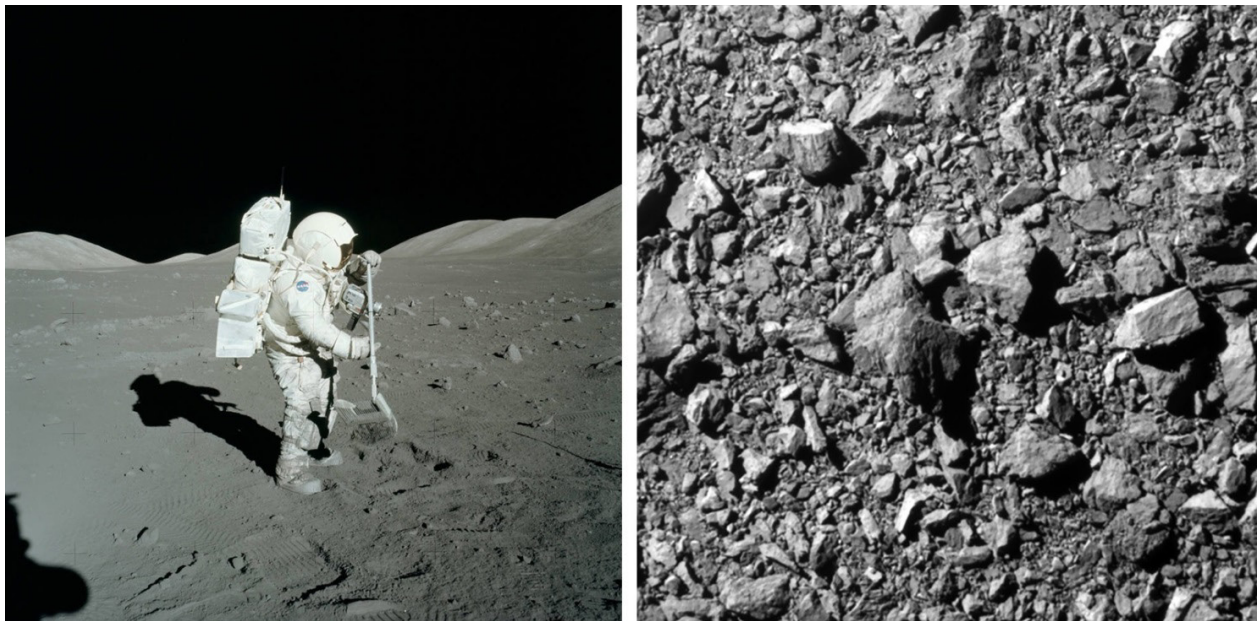

**Figure 4**. The image on the left displays the surface of the Moon, which is coated with a layer of regolith and dust. It captures the moment when NASA astronaut Harrison Schmitt was gathering lunar samples as part of the Apollo 17 mission at the Taurus-Littrow landing site on December 11, 1972. The photo credit goes to Eugene A. Cernan/NASA. On the right side, there is a picture of the rocky surface of the asteroid moonlet Dimorphos, which has a diameter of 150 meters (with a field of view spanning 31 meters across). This image was captured on October 26, 2022, from the DART spacecraft and is credited to NASA/Johns Hopkins APL.

Moon also expands, drawing sustenance from the debris expelled from the Earth's surface.

The distinction between the formation of the Moon and Charon compared to Phobos, Deimos, and binary asteroids lies primarily in the fact that the Moon and Charon are substantial satellites that have moved considerably away from their respective planets due to tidal interactions. In this context, the relatively faster and longer-lasting Moon, characterized by a higher rate of radial drift, likely absorbed the smaller outer satellites of Earth over time, whereas the slower and younger Charon did not have the opportunity to do so. It is conceivable that the extensively discussed enigmatic intense lunar bombardment 3.9 billion years ago, which lacks corresponding evidence on Earth [24], might have resulted from collisions between external smaller satellites and the Moon, which was drifting outward.

It is noteworthy that the transfer of material from a solid-surfaced planet to a satellite disk is absent in the case of gas giant planets. All matter falling from heliocentric orbits onto a gas giant planet becomes incorporated into the planet upon any contact with its surface. Material arriving from heliocentric orbits—both gases and solid bodies—can not only fall onto the planet's surface but can also remain in satellite orbits through fortuitous self-interactions, rather than interactions with the planet itself. This process of gas accretion is extensively elucidated by Ruskol [4]. The typical mass of a satellite system of a gas giant planet is approximately ~$10^{-4}$ of the planet's mass and is found in usually regular and massive satellites, such as Galilean satellites. This gas accretion from heliocentric orbits presumably imparts the planet's rotation rate. An interesting fact emerges: for Jupiter, the equatorial speed of self-rotation is 12.6 km/s, while its orbital speed around the Sun is 13.07 km/s (a difference of 3.7%). This implies that the lower part of Jupiter facing the Sun is almost stationary—similar to the bottom point of a swiftly moving train wheel remaining still relative to the rail. For Saturn, the equatorial rotation speed around its axis is 9.87 km/s, while its orbital speed is 9.68 km/s (a difference of 1.9%). These occurrences are likely not coincidental.

Small and irregular outer satellites of gas giant planets exhibit distinct characteristics from the massive inner satellites. However, their emergence is also linked to quasi-accretionary capture of particles and asteroids from heliocentric orbits. Gorkavyi and Taidakova demonstrated in a series of articles from 1993 to 1995 (references in paper [25]) that the interaction between impacting solid bodies from heliocentric orbits and a protosatellite disk can elucidate the formation of groups of irregular satellites—both prograde, such as the Himalia and Nereid groups, and retrograde, like the Pasiphae and Phoebe groups, along with the massive retrograde Triton. In the 1995 article, a prognostication was put forth concerning the existence of an as-yet-undiscovered outermost cluster of retrograde satellites encircling Saturn. Utilizing numerical simulations involving 256 thousand clusters of debris within the context of the three-body problem, it was determined that the realm occupied by retrograde satellites encompasses approximately 19 to 31 million kilometers, exhibiting a focal concentration of around 25 to 26 million kilometers. The computations indicated that the retrograde satellite Phoebe (positioned at 12.9 million kilometers) should, on the contrary, be predominantly encircled by prograde satellites [26].

In the year 2000, the predicted family of outer retrograde satellites of Saturn was discovered through telescopic observations. By 2023, within the region spanning from 15 to 27 million km, 96 retrograde satellites and only 16 prograde satellites were detected. In the vicinity of Phoebe, ranging from 11 to 15 million km, the reverse trend was observed, with 6 prograde satellites and merely 3 retrograde satellites being identified. Gorkavyi and Taidakova discussed the alignment between theoretical and observational results concerning Saturn's irregular satellites in December 2001. Based on the 1995 model, they inferred an additional outcome that a group of numerous undiscovered satellites likely exists beyond Triton's orbit (> 0.5 million km) around Neptune: "Nereid is probably the largest member of the family of prograde satellites mixed with more numerous family of smaller retrograde satellites" [27]. Observers announced the discovery of such a new Neptune satellite group several months later in 2002. Within the range of 15.7-48.4 million km, 2 prograde and 3 retrograde satellites of Neptune were found. Thus, the formation of satellite systems around gas giant planets is broadly understood and well-described by an accretion model involving interactions between planetesimals from heliocentric orbits and the gaseous protosatellite disk.

## 6. Predictions Regarding Lunar Water

The giant impact theory posits the complete melting of the Moon, followed by its gradual cooling, resulting in a state of near-complete dryness. Jewitt and Young discussed the scenario in which Earth experienced a collision with a Mars-sized object about 4.5 billion years ago. This collision resulted in the ejection of material, which eventually cooled and came together to form the Moon. The immense energy from this colossal impact would have removed a significant portion of Earth's atmosphere, evaporated any existing oceans, and created a deep ocean of molten rock. Whether Earth initially had water or not, the powerful impact that led to the Moon's formation likely removed most, if not all, of the planet's original water [28].

Observations have confirmed the presence of a substantial amount of water on the Moon, prompting the search for mechanisms that deliver new water to the arid lunar surface. Comets appear as the most plausible candidates due to their high ice content and potential encounters within the lunar and terrestrial vicinity. The exact quantity of water brought by comets to the Moon remains uncertain, although their contribution to Earth's oceans has been established. Notably, comets have introduced approximately a quarter of the noble gases, such as xenon, into Earth's atmosphere. Consequently, the contribution of comets to Earth's oceans is estimated to be less than 1% [29]. Furthermore, cometary water exhibits a distinct isotopic composition, with three times the amount of deuterium compared to terrestrial water [30].

He et al. investigate different origins of water on the Moon and perform an examination of the amount, hydrogen isotope makeup, and variations from the core to the rim of water present in impact glass beads obtained from lunar soil samples collected during the Chang'e-5 mission [31]. Their calculations suggest that the amount of water trapped within these impact glass beads within lunar soil could possibly be as high as $2.7 \times 10^{14}$ kilograms [31].

Recently, a model proposing the transfer of water from Earth's atmosphere to the Moon has emerged [32]. The Moon periodically traverses through the Earth's magnetic tail containing terrestrial hydrogen and oxygen ions, suggesting that the Earth's atmosphere could be a contributing source of lunar water. This interaction allows for the accumulation of terrestrial water on the Moon, with a potential capacity of approximately 3000 km$^3$ [32].

Mechanisms related to the solar wind and material transport to polar regions are realistic, but they can only saturate the surface layer with water to a depth of several decimeters or meters. The multi-impact model posits that lunar water shares a common origin with Earth's water—arising from planetesimals that contributed to the formation of both Earth and the Moon. Consequently, water should be present not only in polar but also in equatorial regions of the Moon, where it resides at greater depths. The isotopic composition of lunar water is expected to mirror that of terrestrial water. The surface layers of regolith can contain water from the solar wind, which has a different isotopic composition.

The presence of lunar water should manifest in the lunar geomorphology [33,34]. At the LCROSS impact site (Cabeus crater, approximately 85°S), the concentration of water ice within the regolith is estimated to be 5.6% by mass [35], or potentially even higher [36]. The existence of a permafrost layer beneath the lunar surface can induce alterations in crater shapes, akin to observations in the permafrost zones of Mars [37]. Notably, near certain craters within the Moon's South Pole region, signs of landslides are evident (Figure 5a), exhibiting diverse scale structures [34]—reminiscent of landslides within Earth's permafrost areas (Figure 5b).

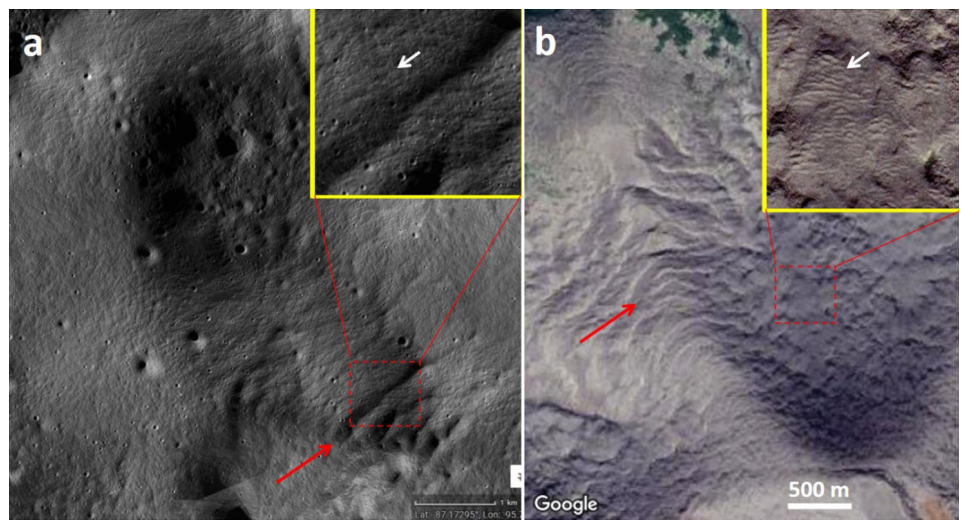

**Figure 5**. (a) The image shows a landslide occurrence along the edge of a crater situated in the region of 87.1°S and 95.0°W on the lunar surface. Notably, visible are cracks spanning many hundreds of meters, as indicated by the red arrow. Additionally, smaller folds measuring tens of meters are denoted by the white arrow (refer to the inset). This figure employs MOONTrek Version: 3.4.140 (source: https://trek.nasa.gov/moon/). (b) This part of the figure displays a patterned ground formation composed of substantial folds, indicated by the red arrow, alongside smaller wrinkles shown by the white arrow. This geomorphic feature occurs within the permafrost melting zone on Earth, situated at coordinates 67.4°N and 134.8°E, and is sourced from Google Maps.

The depth and morphology of craters located at the lunar South Pole may exhibit distinctive characteristics attributed to the presence of a layer of water ice and volatile compounds. These substances are expected to undergo both melting and sublimation processes during impact events and over subsequent periods (refer to Figures 5-6). An examination of photographs depicting the polar regions reveals several notable attributes associated with small craters in proximity to the lunar poles (as illustrated in Figures 5-6):

1) These craters tend to possess smoother contours, and the surrounding terrain displays a distinctive pattern [34].

2) The occurrence of patterned ground is frequently observed within the craters and their immediate vicinity [34].

3) Notably, landslides and cracks are discernible in areas outside the craters [34].

4) Layers or steps are often apparent on the inner slopes of these craters [34].

These characteristics closely resemble those found in craters within the permafrost regions on Mars [38], as well as analogous zones on Earth [39] (refer to Figure 5b). Our hypothesis posits that the observed traits of lunar craters in the South and North Polar regions of the Moon are indicative of the presence of permafrost [34].

When contrasting the topography of the equatorial and polar regions of the Moon, the latter display notably distinctive features (see Figure 6): smoothed craters (Figure 6a) and numerous intricate relief patterns (Figures 6b-6d). It is evident that such features cannot be exclusively attributed to meteoritic impacts. Consequently, an alternative factor is at play, influencing geological activity in the polar zones. This likely involves a substantial layer of perpetual frost, the evolving dynamics of which give rise to the distinct relief patterns known as solifluction structures. The estimated thickness of this layer is on the order of kilometers, as its influence is observable in the shapes of craters of kilometer-scale dimensions (see Figure 6a). Notably, it is improbable for sources of lunar water such as solar wind or Earth's atmospheric tail to create permafrost layers thicker than a few meters.

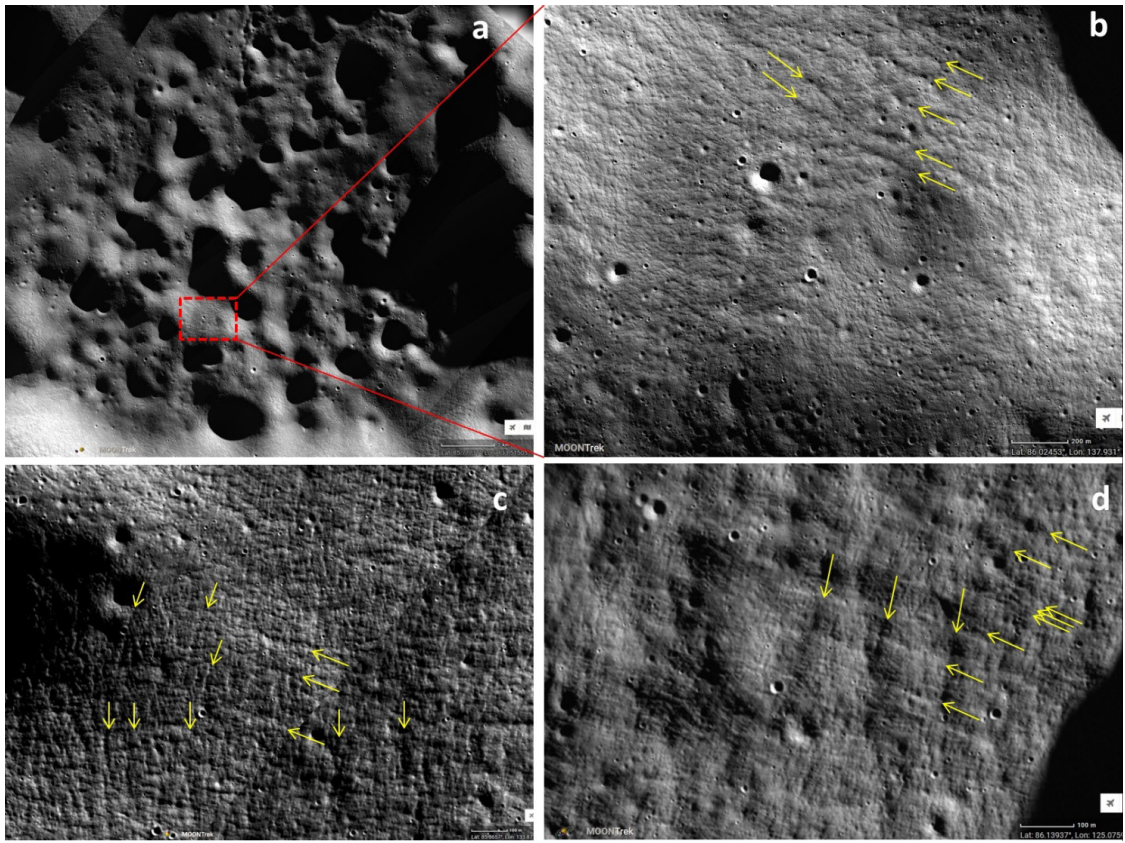

**Figure 6**. (a) Lunar reliefs in the vicinity of the Moon's North Pole exhibit distinctive features, characterized by smoothed forms (the lower left corner of the photograph has coordinates 85.8°N, 133.5°E). (b) Landslide at the edge of a crater (86.0°N, 137.9°E). (c) Patterned surface in the region of 85.9°N, 133.9°E. (d) Regular structures of various scales in the area of 86.1°N, 125.1°E. The yellow arrows in (b, c, d) indicate quasi-regular structures, often with different orientations, which are typical for relief features in permafrost zones [38,39].

Source: NASA/MOONTrek Version: 3.4.140.

## 7. Conclusions

We have demonstrated that ejecta generated by large-scale impacts on a planet's surface readily transition to stable satellite orbits, even when interacting with smaller particles in the protosatellite disk, which are 10-100 times less massive than the ejecta. A protosatellite disk with prograde rotation effectively captures and incorporates ejecta with prograde rotation, destabilizing and returning ejecta with retrograde rotation to the planet. The initial protosatellite disk is not destroyed by the impact of ejecta with a net zero angular momentum, but instead rapidly grows by absorbing prograde ejecta carrying mass and additional angular momentum.

The discussed multi-impact model represents a natural extension of the Hartmann-Davis idea [5] and combines it with the accretion model. The concept of a mega-impact has significantly contributed to understanding lunar formation by emphasizing the importance of a mantle material flow induced by collisions of large planetesimals with the planet. The accretion model explores the formation and evolution of the other crucial element in lunar cosmogony—the protosatellite disk around Earth. These two models—accretion and mega-impact—can be united into a comprehensive multi-impact theory applicable to the formation of a wide spectrum of satellite systems, from large planets like Earth to small asteroids with dimensions of a few hundred meters. From the multi-impact model of lunar formation, significant insights regarding lunar water can be drawn. The mega-impact theory would have completely vaporized water from lunar rocks and necessitated the introduction of new mechanisms for lunar water formation—involving comets, solar wind particles, or Earth's atmosphere. Such mechanisms imply shallow subsurface lunar ice in relatively small quantities. The multi-impact model posits that the Moon possesses substantial water reserves from planetesimals that also contributed to Earth's formation. The material from Earth's crust that contributed to the Moon likely held a considerable amount of water as well. It is plausible that lunar water is distributed beneath the surface at all latitudes, albeit at different depths. This can lead to the formation of regular geomorphological structures, particularly in polar regions, (see, for example, [33,34,38,39]). Determining the isotopic composition of lunar water will provide a definitive answer about its origin and the most probable scenario of lunar formation.

## Funding

This research received no external funding.

## Acknowledgement

The author expresses gratitude to Tanya Taidakova for years of fruitful collaboration; John Mather for support and valuable advice, and astronaut Harrison Schmitt for his insightful lecture at the Goddard Space Flight Center in May 2017.

## Data Availability Statement

The data supporting these findings are publicly available.

## Conflict of Interest

The author declares no conflicts of interest.